\documentclass[12pt,authoryear,3p,times]{elsarticle}
\usepackage{setspace}
\usepackage{amssymb}
\usepackage{amsmath}
\usepackage{bbm}
\usepackage{enumitem}
\usepackage{booktabs}
\usepackage{hyperref}
\usepackage[utf8]{inputenc}
\usepackage[T1]{fontenc} 
\usepackage{multirow}
\usepackage[table,xcdraw]{xcolor}
\usepackage[dvipsnames]{xcolor} 
\usepackage{float}

\usepackage{multicol}
\usepackage{orcidlink}

\journal{EJOR}

\begin{document}

\begin{frontmatter}

\title{Probabilistic forecasting via post-processing prediction errors:\\In- or out-of-sample?}
\author[ORBI]{Piotr Zaborowski \orcidlink{0009-0006-1245-6568}}
\ead{piotr.zaborowski@pwr.edu.pl}
\author[ONAS]{Arkadiusz Lipiecki \orcidlink{0000-0003-1118-0388}}
\ead{arkadiusz.lipiecki@pwr.edu.pl}
\author[BATH]{Fotios Petropoulos \orcidlink{0000-0003-3039-4955}\corref{cor1}}
\cortext[cor1]{Corresponding author}
\ead{f.petropoulos@bath.ac.uk}
\author[ORBI]{Rafał Weron \orcidlink{0000-0003-1619-5239}}
\ead{rafal.weron@pwr.edu.pl}
\address[ORBI]{Department of Operations Research and Business Intelligence, Wrocław University of Science and Technology, Poland}
\address[ONAS]{Department of Computational Social Science, Wrocław University of Science and Technology, Poland}
\address[BATH]{School of Management, University of Bath, United Kingdom}

\begin{abstract}

Many forecasting systems produce point forecasts even when decisions require information about uncertainty. We investigate whether post-processing methods can systematically improve upon traditional Gaussian predictive distributions constructed from in-sample residuals. We propose a hybrid framework that combines forecast error post-processing with model-specific scaling of forecast uncertainty across horizons. For a comprehensive evaluation, we apply historical simulation, conformal prediction, quantile regression, and GARCH-based post-processing to point forecasts generated by Theta, exponential smoothing, and ARIMA models. Using 14,407 monthly series from the M4 competition and forecast horizons of 1 to 12 months, we evaluate performance using the continuous ranked probability score and rank-based statistical comparisons. Averaged across horizons, all post-processing variants improve upon the benchmark predictive distributions, with gains of up to 4.6\%. In-sample calibration outperforms its out-of-sample counterpart in 11 of the 12 model--method combinations, although the preferred post-processing method depends on the base model and forecast horizon. The advantage of in-sample calibration generally increases at longer horizons. Our results show that organisations can extend existing point-forecasting systems to provide useful uncertainty quantification without computationally intensive repeated model re-estimation.
\end{abstract}


\end{frontmatter}

\newpage

\begin{multicols}{2}
{\scriptsize
\tableofcontents    
}
\end{multicols}

\section{Introduction}
\label{sec:Introduction}

Forecasts are an indispensable component of decision making. Virtually all decisions rely on forecasts, whether these forecasts are implicit or explicit \citep{Makridakis2024-nr,Petropoulos2022-xr}. Although forecasts are sometimes embedded within processes or algorithms in which the decision itself is the primary output, as in prescriptive analytics applications, they are more commonly used explicitly (and revised when appropriate) to support decision making. In many cases, only \textit{point forecasts}, i.e., single-number estimates, are provided or used. Point forecasts are easier to understand and communicate and often align well with operational constraints and existing key performance indicators \citep{Gneiting2014-zs,Goodwin2014-jv,Raftery2016-zf}. However, they can create an illusion of precision because they do not directly convey the uncertainty surrounding future outcomes. When risk is important, extreme outcomes are costly, decisions are asymmetric, or scenario planning is required, \textit{probabilistic forecasts} that explicitly represent uncertainty are preferable \citep{Taylor2026-ly}. Such settings arise, for example, in weather forecasting, financial risk management, epidemiology, energy forecasting, and supply chain forecasting \citep{Jeon2019-ev,lip:uni:wer:24,Taylor2023-ut,Wang2024-fo}.

Increasingly, forecasting software and packages provide uncertainty estimates alongside point forecasts \citep{forecast2,lip:wer:25}. These may take the form of prediction intervals, which specify a range expected to contain a future observation with a given probability, quantile forecasts, which estimate selected quantiles of the predictive distribution, or full predictive distributions. However, such uncertainty estimates often rely on simplifying assumptions \citep{Chatfield1993-gn,hyn:ath:21:3ed}, including correct model specification and parameter estimation, as well as normality, homoskedasticity, and independence of forecast errors. Traditionally, uncertainty estimates are often based on in-sample one-step-ahead residuals, with their uncertainty propagated across forecast horizons according to model-specific formulas. Particular model classes may impose additional assumptions, e.g., ARIMA models require the series to be stationary after appropriate differencing and/or transformations. When these assumptions are violated, the resulting probabilistic forecasts may be unreliable and, in particular, may underestimate forecast uncertainty.

An alternative to analytical formulae for deriving (theoretical) prediction intervals is to post-process forecast errors and use their empirical distribution to quantify forecast uncertainty. In practice, post-processing provides a way to translate point forecasts into probabilistic forecasts. The literature has proposed several post-processing approaches that differ in their assumptions and in how they use past forecast errors \citep{lip:uni:wer:24,sha:vov:08,van:etal:21:postprocessing}:
\begin{itemize}[noitemsep]
\item historical simulation (HS) and conformal prediction (CP) construct probabilistic forecasts by adjusting point forecasts using empirical quantiles of past forecast errors; HS uses signed errors, whereas CP typically uses absolute errors as non-conformity scores to construct symmetric prediction intervals;
\item quantile regression (QR) estimates conditional quantiles as functions of point forecasts by fitting quantile-specific regression coefficients;
\item generalized autoregressive conditional heteroskedasticity (GARCH) models time-varying forecast-error variance using lagged conditional variances and squared errors, and converts the resulting conditional variance forecasts into probabilistic forecasts under an assumed error distribution.
\end{itemize}

This paper makes three contributions. First, we present a hybrid framework for post-processing point forecasts that can be calibrated using either in-sample residuals or out-of-sample forecast errors. The framework combines post-processing methods with model-based forecast-error variance estimates to (i) adjust the scale of in-sample residuals to reflect out-of-sample forecast uncertainty and (ii) propagate uncertainty across longer forecast horizons. We implement and compare the in-sample and out-of-sample calibration approaches using four post-processing methods: HS, CP, QR, and GARCH. 
Although in-sample errors are readily available at negligible additional computational cost, their use in forecast post-processing has received comparatively little attention and, to the best of our knowledge, has been considered mainly for GARCH, whereas other post-processing approaches rely on out-of-sample errors. 
Second, we conduct a large-scale evaluation using 14,407 heterogeneous monthly series from the M4 competition and three commonly used forecasting models: Theta, ETS, and ARIMA. Third, we examine how relative forecast performance varies across horizons and quantify the computational costs of in-sample and out-of-sample calibration.

We use a diverse set of real-world time series to evaluate whether post-processing point forecasts can improve probabilistic forecasts relative to the standard uncertainty estimates provided by widely used open-source forecasting implementations, such as the \texttt{forecast} package for R \citep{forecast2}. To construct probabilistic forecasts from prediction errors, we use the \texttt{PostForecasts.jl} package for Julia \citep{lip:wer:25}. Forecast performance is evaluated using the continuous ranked probability score (CRPS), a proper scoring rule that assesses the quality of the predictive distribution \citep{gne:raf:07}. We examine performance across forecast horizons and assess statistical significance using multiple comparisons with the best (MCB) tests \citep{Koning2005-iq}. Our results show that post-processing prediction errors can significantly improve probabilistic forecast performance, with the gains generally increasing at longer horizons.

The rest of the paper is organized as follows. Section~\ref{sec:post-processing} describes the post-processing methods used to construct probabilistic forecasts from prediction errors. Section~\ref{sec:design} presents the empirical design, including the data, forecasting models, and forecast evaluation measures. Section~\ref{sec:Results} reports the empirical results and statistical significance analysis. Section~\ref{sec:Discussion} discusses the findings, with particular emphasis on their implications and limitations. Finally, Section~\ref{sec:Conclusions} summarizes the main findings and outlines directions for future research.

\section{Post-processing prediction errors}
\label{sec:post-processing}

\subsection{Notation}
\label{ssec:Notation}

Let us first introduce the notation used throughout the paper. We denote time series $i$ by $\{y_{i,t}\}_{t=1}^{T}$, where $i=1,\ldots,N$. The final $K$ observations of each series form the test period,
$\mathcal{T}_{\text{test}}=\{T-K+1,\ldots,T\}$.
We use $\tau$ to denote the \textit{target} time point for which a forecast is produced. For an $h$-step-ahead forecast, the corresponding \textit{forecast origin} is $\xi=\tau-h$.

A point forecast $\hat{y}_{i,\tau|\xi}$ is obtained by fitting a forecasting model $f(\cdot)$ to the observations available at the forecast origin: $\hat{y}_{i,\tau|\xi} = f\left(y_{i,1},\ldots,y_{i,\xi};\tau\right)$. We refer to the observations used to fit $f(\cdot)$ as the \textit{training sample} and to the model-fitting process as \textit{training}.

Similarly, a quantile forecast $\hat{q}_{i,\tau|\xi}^{p}$ for quantile level $p\in(0,1)$ can be obtained analytically (see also Section \ref{ssec:Point:Models}) or by post-processing the point forecasts and corresponding realized observations available at the forecast origin $\xi$. We refer to the set of predictions and observations used for post-processing as the \textit{calibration sample}, with $\mathcal{C}_{\xi}$ denoting the set of its time indices, and to the process itself as \textit{calibration}. 

\subsection{Historical simulation and conformal prediction}
\label{ssec:HS:CP}

Historical simulation \cite[HS;][]{ale:08,now:wer:18} is a model-agnostic approach that constructs probabilistic forecasts by combining a point forecast $\hat{y}_{i,\tau|\xi}$ with the empirical distribution of prediction errors from the calibration sample $\mathcal{C}_{\xi}$. The quantile forecast for probability level $p\in(0,1)$ is given by
\begin{equation}
\label{eq:HS}
\hat{q}^{p}_{i,\tau|\xi} = \hat{y}_{i,\tau|\xi} +
Q_p\left(\{\varepsilon_{i,t}\}_{t\in\mathcal{C}_{\xi}}\right),
\end{equation}
where $Q_p(\cdot)$ denotes the sample $p$-quantile, computed according to Definition 7~of~\cite{hyn:fan:96}, and $\varepsilon_{i,t}=y_{i,t}-\hat{y}_{i,t}$ denotes the prediction error associated with observation $t\in\mathcal{C}_{\xi}$. The term \textit{historical simulation} can be traced back to the early 1990s and the development of Value-at-Risk estimation in financial risk management \citep{hen:96}, although the use of empirical prediction-error distributions for probabilistic forecasting was proposed earlier \citep{wil:goo:71}.

The related concept of \textit{conformal prediction} (CP) originated in the machine-learning literature \citep{vov:gam:sha:05}, although the use of absolute errors to construct prediction intervals had already been discussed by \cite{wil:goo:71}. Like HS, CP is model-agnostic and relies on the empirical distribution of prediction errors \citep{kat:zie:21}. The main difference is that classical CP typically produces a \textit{prediction interval} (PI), which is symmetric around the point forecast when absolute errors $|\varepsilon_{i,t}|=|y_{i,t}-\hat{y}_{i,t}|$ are used as non-conformity scores $\lambda_{i,t}$. The latter are measures of how unusual or poorly predicted an observation is relative to a fitted model and the other data. 

The quantile forecast for probability level $p\in(0,1)$ can be obtained by shifting the point forecast by an empirical quantile of the non-conformity score distribution:
\begin{equation}\label{eq:CP1}
    \hat{q}^p_{i,\tau|\xi} = \hat{y}_{i,t|\xi} - \mathbbm{1}_{p \le 0.5} Q_{2p}(\{\lambda_{i,t}\}_{t\in\mathcal{C}_{\xi}}) + \mathbbm{1}_{p \ge 0.5} Q_{2(1 - p)}(\{\lambda_{i,t}\}_{t\in\mathcal{C}_{\xi}}).    
\end{equation}
While $\hat{y}_{i,t|\xi} \pm Q_{p}(\{\lambda_{i,t}\}_{t\in\mathcal{C}_{\xi}})$ is a valid prediction interval of nominal coverage $p$ without any additional assumptions other than exchangeability, translating it to quantiles can be performed only under the assumption of a symmetric error distribution. HS can be considered a variant of CP, which uses signed prediction errors, rather than their absolute values. Hence, for both methods, we use the \texttt{CP} model from the PostForecasts.jl package in Julia~\citep{lip:wer:25}, with the keyword argument \texttt{abs} set to \texttt{true} for CP and \texttt{false} for HS.

\subsection{Quantile regression}
\label{ssec:QR}

Quantile regression models a specified conditional quantile of a response variable as a function of predictors \citep{koe:17}. It can be used for post-processing by estimating that quantile from point forecasts and the corresponding observed values. Quantile Regression Averaging (QRA), introduced by \citet{now:wer:15} and successfully applied in the GEFCom2014 forecasting competition \citep{hon:pin:fan:etal:16,mac:now:16}, is a widely used post-processing method in energy forecasting \citep{liu:now:hon:wer:17,now:wer:18,wan:etal:19,yan:yan:liu:23,cor:din:pou:24,mac:ser:uni:24}. Its original formulation constructs quantile forecasts as linear combinations of a pool of point forecasts. If the forecasts in the pool are first averaged and the resulting average is then used as the sole regressor in QR, the method is referred to as Quantile Regression Machine \cite[QRM;][]{mar:uni:wer:20IJF,uni:23}. 

In this study, we use a single point forecast as the regressor, so QRA reduces to the QRM specification; for notational simplicity, we refer to this post-processing method as \textit{quantile regression} (QR). The quantile forecast conditional on the point forecast is given by
\begin{equation}
\label{eq:QRA}
\hat{q}^{p}_{i,\tau|\xi} = \beta_{1,p}\hat{y}_{i,\tau|\xi} + \beta_{0,p},
\end{equation}
where $\beta_{0,p}$ and $\beta_{1,p}$ are estimated separately for each quantile level $p\in(0,1)$ using the linear programming formulation \citep{koe:17}. Because the models are estimated independently across quantile levels, the resulting quantile forecasts need not be non-decreasing in $p$, a problem commonly known as quantile crossing \citep{koe:bas:82,che:f-v:gal:10}. When quantile crossing occurs, we sort the predicted quantiles to obtain a non-decreasing sequence \citep{now:wer:15}. We use the \texttt{QR} model from the PostForecasts.jl package~\citep{lip:wer:25}, which solves the optimization problem using HiGHS~\citep{hua:hal:18}.

\subsection{GARCH}
\label{ssec:GARCH}

Generalized autoregressive conditional heteroskedasticity \cite[GARCH;][]{bol:1986} is a seminal framework for modeling time-varying volatility in financial markets. In this paper, we consider the widely used GARCH(1,1) specification,
\begin{equation}
\label{eq:garch}
\hat{\sigma}^2_{i,\tau|\xi}
=
\omega
+
\alpha\varepsilon^2_{i,\tau-1}
+
\beta\hat{\sigma}^2_{i,\tau-1|\xi},
\end{equation}
where $\hat{\sigma}^2_{i,\tau|\xi}$ denotes the conditional prediction error variance for series $i$ at time $\tau$, given the information available at forecast origin $\xi$. The parameters are estimated by maximum likelihood under the assumption that the standardized forecast errors are normally distributed, $\varepsilon_t \sim \mathcal{N}(0, \hat{\sigma}_t)$, using the method of moving asymptotes (MMA) \citep{sva:02}. For this task we use the \texttt{GARCH} model from the PostForecasts.jl package~\citep{lip:wer:25}, with the NLopt~\citep{NLopt} implementation of the MMA algorithm.

The intercept $\omega$ is determined using variance targeting, which can reduce computational cost and improve estimation robustness under model misspecification \citep{fra:etal:11}. Specifically,
$\omega=\bar{\varsigma}^2_{i,\xi}(1-\alpha-\beta)$,
where $\bar{\varsigma}^2_{i,\xi}$ is the sample variance of the forecast errors in the calibration window available at origin $\xi$. This formulation requires $\alpha+\beta<1$ to ensure covariance stationarity and a finite unconditional variance.

For multi-step-ahead volatility forecasting, the future error $\varepsilon_{i,\tau-1}$ is unknown. Its squared value in Eq.~\eqref{eq:garch} is therefore replaced by its conditional expectation,
$\mathbb{E}_{\xi}[\varepsilon^2_{i,\tau-1}]
=
\hat{\sigma}^2_{i,\tau-1|\xi}$,
which yields
\begin{equation}
\label{eq:garch:multi}
\hat{\sigma}^2_{i,\tau|\xi}
=
\omega
+
(\alpha+\beta)\hat{\sigma}^2_{i,\tau-1|\xi}.
\end{equation}
Assuming conditionally Gaussian forecast errors, the quantile forecast for target time $\tau$ and quantile level $p$ is
\begin{equation}
\label{eq:garch:q}
\hat{q}^{p}_{i,\tau|\xi}
=
\hat{y}_{i,\tau|\xi}
+
\hat{\sigma}_{i,\tau|\xi}
\sqrt{2}\,\text{erf}^{-1}(2p-1),
\end{equation}
where $\sqrt{2}\,\text{erf}^{-1}(2p-1)$ is the $p$-quantile of the standard normal distribution.

\subsection{In-sample vs.\ out-of-sample post-processing}
\label{ssec:in:out:post-processing}

\citet{mak:win:89} distinguished between two types of forecast errors: in-sample residuals and out-of-sample forecast errors, emphasizing that the latter are more relevant for forecasting applications. Post-processing based on out-of-sample errors typically requires splitting the data into training and calibration or validation samples, repeatedly estimating model parameters, for example through cross-validation, or maintaining an archive of historical forecasts and corresponding realizations. By contrast, in-sample methods construct probabilistic forecasts directly from residuals obtained when fitting the model. This approach is computationally simpler, but it relies on the strong assumption that the distribution of in-sample residuals provides an adequate approximation to the distribution of out-of-sample forecast errors.

Among the post-processing methods considered in this study, CP \citep{sha:vov:08,kat:zie:21,zaf:etal:22} and QR \citep{now:wer:15,uni:22,lip:uni:wer:24} are typically, although not exclusively \citep{tay:bun:99}, applied to out-of-sample forecasts or forecast errors. For HS, to the best of our knowledge, \citet{wil:goo:71} were the first to use the empirical distribution of out-of-sample forecast errors to construct prediction intervals. Variants of this approach have subsequently been applied in a range of forecasting settings \citep{sil:mou:00,pin:kar:10,kaa:etal:17,lip:uni:wer:24}. However, the empirical distribution of model residuals has also been applied to produce out-of-sample probabilistic forecasts~\citep{tay:21, tay:men:26}. By contrast, GARCH models are typically estimated in-sample, either jointly with the parameters of the conditional mean model \citep{bol:1986,zha:07, tay:jeo:18, tay:men:26} or in a two-step procedure in which the GARCH parameters are estimated from the residuals of the mean model \citep{jan:puc:23,leb:etal:26}. By evaluating all four methods under both in-sample calibration, denoted by $\text{HS}_{\text{in}}$, $\text{CP}_{\text{in}}$, $\text{QR}_{\text{in}}$, and $\text{GARCH}_{\text{in}}$, and out-of-sample calibration, denoted by $\text{HS}_{\text{out}}$, $\text{CP}_{\text{out}}$, $\text{QR}_{\text{out}}$, and $\text{GARCH}_{\text{out}}$, we aim to provide new evidence on their relative forecasting performance.

\subsection{Horizon-specific scale adjustment}
\label{ssec:Scale:adjustment}

The forecasting models used in this study include established procedures for constructing PIs under the assumption of normally distributed forecast errors; see Section \ref{ssec:Point:Models} for details. These procedures adjust the residual variance to account for parameter estimation, through a degrees-of-freedom correction, and for the forecast horizon. Because the resulting model-based predictive distributions serve as benchmarks for the post-processing methods, we use the corresponding horizon-specific scale adjustments to convert post-processed one-step-ahead distributions into multi-step-ahead predictive distributions.

This yields a hybrid approach that combines the post-processing methods described in Sections \ref{ssec:HS:CP}--\ref{ssec:GARCH} with horizon-dependent uncertainty estimates obtained from the fitted forecasting models. For series $i$, forecast origin $\xi$, horizon $h$, and target $\tau=\xi+h$, the base model provides the forecast-error standard deviation $\hat{\varsigma}_{i,\tau|\xi}$. This quantity is derived from the model fitted at origin $\xi$ and is described in detail in Section~\ref{ssec:Point:Models}. The model-based horizon scaling is applied to both the in-sample and out-of-sample post-processing variants, with some modifications.

In the in-sample post-processing approach, we assume that the distribution of the model residuals provides useful information about the shape of the future forecast-error distribution, while its scale varies with the forecast horizon according to the underlying forecasting model. Under this scheme, the quantile forecasts obtained from a post-processing method, $\hat{q}^{p}_{i,\tau|\xi}$, are rescaled for each horizon $h$ as follows:
\begin{equation}
\tilde{q}^{p}_{i,\tau|\xi} = \hat{q}^{0.5}_{i,\tau|\xi} + \left(\hat{q}^{p}_{i,\tau|\xi} - \hat{q}^{0.5}_{i,\tau|\xi} \right) \frac{\hat{\varsigma}_{i,\tau|\xi}}{\bar{\varsigma}^{\text{in}}_{i,\xi}}, \label{eq:post-in-sample}
\end{equation}
where $\tilde{q}^{p}_{i,\tau|\xi}$ is the final forecast of the $p$-quantile for series $i$ and target $\tau=\xi+h$, $\hat{\varsigma}_{i,\tau|\xi}$ is the model-specific forecast-error standard deviation at horizon $h$, and $\bar{\varsigma}^{\text{in}}_{i,\xi}$ is the sample standard deviation of the model residuals available at origin $\xi$. The transformation preserves the median of the post-processed distribution while adjusting its dispersion according to the horizon-specific uncertainty estimate provided by the forecasting model. It therefore allows the shape of the predictive distribution to be estimated without imposing normality.

In the out-of-sample post-processing approach, the quantile forecasts are rescaled in the same way, but only for horizons $h>1$. Because the calibration sample consists of one-step-ahead out-of-sample forecast errors, no additional scale adjustment is required at $h=1$ and $\tilde{q}^{p}_{i,\xi+1|\xi} = \hat{q}^{p}_{i,\xi+1|\xi}$. For $h=2,\ldots,H$, the final quantile forecast is given by
\begin{equation}
\tilde{q}^{p}_{i,\tau|\xi} = \hat{q}^{0.5}_{i,\tau|\xi} + \left( \hat{q}^{p}_{i,\tau|\xi} - \hat{q}^{0.5}_{i,\tau|\xi} \right) \frac{\hat{\varsigma}_{i,\tau|\xi}} {\hat{\varsigma}_{i,\xi+1|\xi}}. 
\label{eq:post-out-of-sample}
\end{equation}
Thus, the empirical distribution of one-step-ahead out-of-sample errors determines the shape and initial scale of the predictive distribution, while the ratio of model-specific forecast-error standard deviations determines how its dispersion changes with the horizon.

\section{Experimental design}
\label{sec:design}

\subsection{Data}
\label{ssec:Data}

Forecasting competitions have been fertile playgrounds in terms of open-access, publicly available real-life data for testing and evaluating new research ideas. One of the most important competitions of the last ten years is the M4 forecasting competition \citep{Makridakis2018-oy,Makridakis2020-mm}. The M4 data consist of 100,000 time series in total, across different data frequencies: yearly, quarterly, monthly, weekly, daily, and hourly. Data in the M4 competition come from different domains, including macro, micro, demographic, industry, finance, and other. \cite{Spiliotis2020-gl} demonstrated that forecasting competition data, such as the M4 data, are representative pools of data with regards to statistical time series features, and concluded that ``M4 could become a standard testing ground for evaluating the performances of generic time series methods''. 

In this study, we focus on the monthly M4 data set, which consists of 48,000 time series. From those, we use the 14,411 longest time series, for which there are at least $T=324$ available observations (i.e., 27 years of data). The reason for using only the longest time series simply relies on the need for a decent-sized calibration window to complete the out-of-sample post-processing of the prediction errors in order to derive probabilistic distributions for the methods described in Section \ref{sec:post-processing}. To keep the study design simple, we trimmed the time series that were longer than $T=324$ by dropping the corresponding number of initial observations. Finally, we removed four series whose observations remained unchanged throughout the final six years, i.e., the last 72 months, leaving a final sample of $N=14{,}407$ time series. For context, the included time series used in this study are reasonably balanced across the M4 data categories, see Table \ref{tab:datacategories}.

\begin{table}[tb]
\caption{The distribution of the 14,407 selected series across the different categories of the M4 forecasting competition.}
\small
\centering
\begin{tabular}{cccccc}
\hline
Macro & Micro & Demographic & Industry & Finance & Other \\
\hline
3,818 & 3,416 & 3,159 & 2,333 & 1,634 & 47 \\
\hline
\end{tabular}
\label{tab:datacategories}
\end{table}

\subsection{Training and calibration}
\label{ssec:Training}

The final $K=12$ observations of each series form the test period,
$\mathcal{T}_{\text{test}}=\{313,\ldots,324\}$. Although this is a relatively short test window for an individual series, the large cross-sectional dimension of the data supports the robustness of the aggregate empirical results. Overall, the test sample forms a $K\times N=12\times14{,}407$ panel containing $172{,}884$ monthly observations. Each observation is evaluated at all $H=12$ forecast horizons, $h=1,\ldots,12$, resulting in $K\times N\times H=2{,}074{,}608$ forecast--observation evaluations for each forecasting and post-processing method.

\begin{figure}[tb]
    \centering
    \includegraphics[width=1\linewidth]{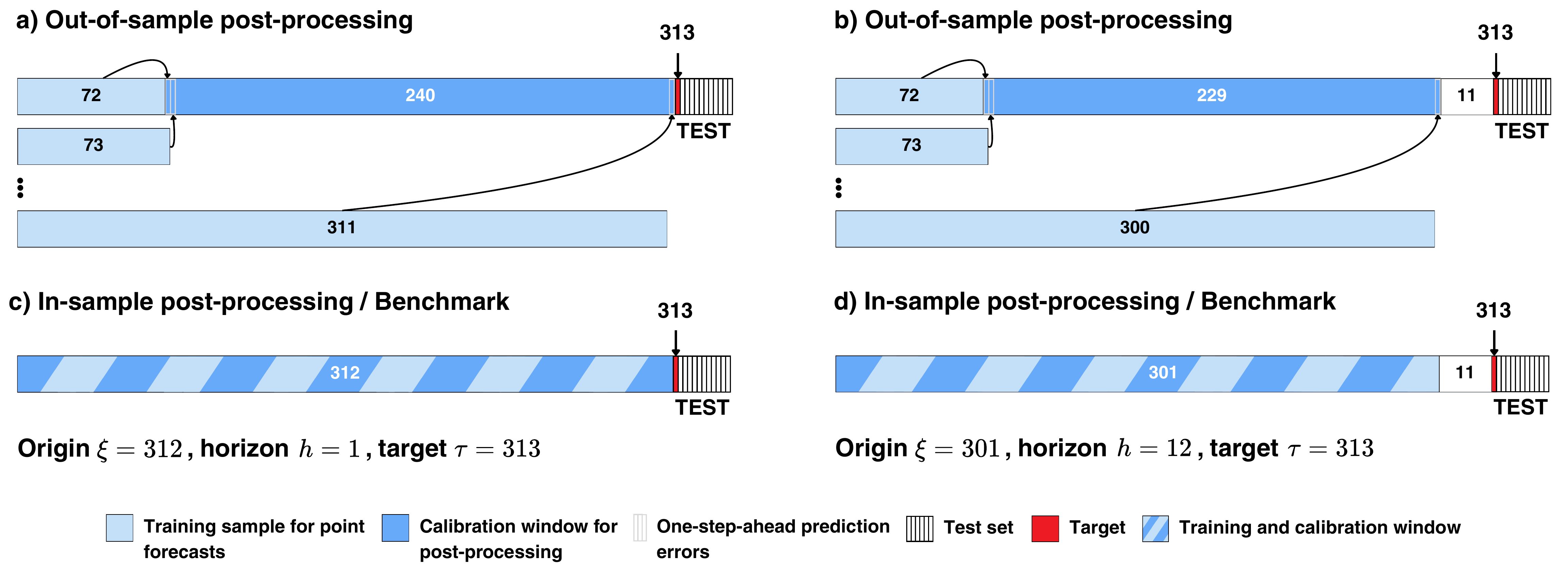}
    \caption{Illustration of the training data and forecasts used for calibration and evaluation in two settings: forecast origin $\xi=312$, horizon $h=1$, and target $\tau=313$ in panels a) and c); and forecast origin $\xi=301$, horizon $h=12$, and target $\tau=313$ in panels b) and d). In panels a) and b), the light blue and dark blue bars indicate the training and calibration samples, respectively, for out-of-sample post-processing. In panels c) and d), the hatched light blue and dark blue bars indicate the observations used jointly for training and in-sample calibration. In all panels, the white bars indicate unused observations, while the outlined rectangles mark the test period $\mathcal{T}_{\text{test}}=\{313,\ldots,324\}$, with the target observation highlighted in red.}
    \label{fig:scheme}
\end{figure}

Point forecasts $\hat{y}_{i,\tau|\xi}$ are generated using an expanding-window scheme. We consider forecast origins $\xi=72,\ldots,323$ and horizons $h=1,\ldots,12$, where the forecast target is $\tau=\xi+h$; see Figure~\ref{fig:scheme}. Among the forecasts with targets $\tau\leq312$, only the one-step-ahead forecasts are retained for out-of-sample calibration, as described in Section~\ref{ssec:in:out:post-processing}. Forecasts for horizons $h>1$ with targets $\tau\leq312$ and forecasts with targets beyond the end of the observed series, $\tau>T=324$, are discarded. Forecasts with targets $\tau\in\mathcal{T}_{\text{test}}$ are converted into quantile forecasts using the in-sample and out-of-sample post-processing procedures defined in Eqs.~\eqref{eq:post-in-sample} and~\eqref{eq:post-out-of-sample}, respectively.

For the out-of-sample post-processing methods $\text{HS}_{\text{out}}$, $\text{CP}_{\text{out}}$, $\text{QR}_{\text{out}}$, and $\text{GARCH}_{\text{out}}$, the quantile forecast $\hat{q}_{i,\tau|\xi}^{p}$ is calibrated using the realised observations and the corresponding out-of-sample one-step-ahead forecasts: $\{(y_{i,t},\hat{y}_{i,t|t-1})\}_{t=73}^{\xi}$, as illustrated in Figure~\ref{fig:scheme}, panels a) and b) for target $\tau=313$ and two forecast origins $\xi=312$ and $301$, respectively. 

For the in-sample post-processing methods $\text{HS}_{\text{in}}$, $\text{CP}_{\text{in}}$, $\text{QR}_{\text{in}}$, and $\text{GARCH}_{\text{in}}$, the quantile forecast $\hat{q}_{i,\tau|\xi}^{p}$ is calibrated using the observations and the corresponding fitted values from the forecasting model trained at origin $\xi$: $\{(y_{i,t},\hat{y}_{i,t|\xi})\}_{t=1}^{\xi}$, as illustrated in Figure~\ref{fig:scheme}, panels c) and d) for target $\tau=313$ and two forecast origins $\xi=312$ and $301$, respectively. 

Note the subtle difference in the length of the calibration samples. In out-of-sample calibration, each forecast $\hat{y}_{i,t|t-1}$ is generated by a model fitted at a different origin, namely $t-1=72,\ldots,\xi-1$. By contrast, all values $\hat{y}_{i,t|\xi}$ used for in-sample calibration are obtained from the model fitted at the current origin $\xi$. Hence, the out-of-sample calibration sample contains 72 forecast--observation pairs fewer than the in-sample calibration sample. 

\subsection{Forecasting models}
\label{ssec:Point:Models}

We consider three univariate forecasting models: the Theta method, exponential smoothing, and ARIMA. These models are widely used as benchmarks in the forecasting literature because they are straightforward to implement, computationally efficient, and robust across a wide range of time-series patterns \citep{Makridakis2020-mm,Petropoulos2022-xr}. We use the \texttt{forecast} package for R to generate forecasts from all three models \citep{forecast2}.

\subsubsection{Theta} 

The Theta method is a well-established univariate forecasting approach introduced by \citet{As:Nik:00}. Despite its structural simplicity, it achieved the best performance in the M3 competition \citep{Mak:Hib:00}. In the M4 competition, it continued to exhibit strong forecast accuracy, although it was surpassed by more sophisticated hybrid methods \citep{Makridakis2020-mm}.

We use the \texttt{thetaf()} function from the \texttt{forecast} package in R. For each model fit, forecasts are generated for horizons $h=1,\ldots,12$ months. Prior to training, each series is automatically tested for seasonality. If seasonal patterns are detected using an autocorrelation-based test at lag $H=12$, the series is adjusted using classical multiplicative decomposition, and the resulting forecasts are deseasonalized.

Following \citet{Hyn:Bil:03}, the point forecast at horizon $h$ is given by
\begin{equation}
\hat{y}_{t+h} = \tilde{y}_{t+h} + \frac{\hat{b}}{2} \left( h-1+ \frac{1-(1-\alpha)^\xi}{\alpha} \right),
\end{equation}
where $\tilde{y}_{t+h}$ is the $h$-step-ahead simple exponential smoothing (SES) forecast, $\hat{b}$ is the slope of the linear trend fitted to the training sample, $\alpha$ is the SES smoothing parameter, and $\xi$ is the length of the training sample.

Prediction intervals are obtained analytically from the underlying state space model \citep{Hyn:Bil:03}. The standard error of the forecast at horizon $h$ is given by
\begin{equation}
\hat{\varsigma}_h = \tilde{\sigma} \sqrt{1+(h-1)\alpha^2},\end{equation}
where $\tilde{\sigma}^2$ is the residual variance estimated from the SES model fitted to the deseasonalized series. Prediction intervals are then constructed using standard normal quantiles under the assumption of Gaussian forecast errors.

\subsubsection{Exponential smoothing}

Exponential smoothing (ETS) is a family of statistical methods designed for time series forecasting. ETS models are widely used in applied forecasting and have demonstrated strong performance in the M3 and M4 competitions \citep{Mak:Hib:00,Makridakis2020-mm}.

We use the \texttt{ets()} function from the \texttt{forecast} package in R. By default, it considers 15 exponential smoothing models, excluding specifications with multiplicative trends as well as combinations of components that may lead to numerical instability. The function selects the ``best'' model out of the 15 based on the Akaike's Information Criterion corrected for small sample sizes (AICc). 

Under the default settings of \texttt{ets()}, prediction intervals are obtained from the fitted ETS state space model. The $h$-step-ahead forecast-error variance is derived by propagating future innovation uncertainty through the state equations, using analytical results where available and linearization approximations for nonlinear multiplicative models. Assuming Gaussian forecast errors, the corresponding prediction interval is constructed as
\begin{equation}
\hat{q}^{p}_{\tau|\xi}
=
\hat{y}_{\tau|\xi}
+
z_p\,\hat{\varsigma}_{\tau|\xi},
\qquad
z_p = \Phi^{-1}(p),
\label{eq:ETS:PI}
\end{equation}
where $z_p = \Phi^{-1}(p)$ denotes the $p$-quantile of the standard normal distribution and $\hat{\varsigma}_{\tau|\xi}$ is the estimated standard deviation of the forecast error at target $\tau = \xi+h$. Thus, the predictive distribution is Gaussian with mean $\hat{y}_{\tau|\xi}$ and horizon-specific standard deviation $\hat{\varsigma}_{\tau|\xi}$.

\subsubsection{ARIMA}

ARIMA is a classical family of statistical models for time series analysis and forecasting, introduced by \cite{box:jen:70}. ARIMA models temporal dependence using autoregressive (AR), differencing (I), and moving average (MA) components, characterized by the orders $(p,d,q)$. The family also includes SARIMA (Seasonal ARIMA) models, which incorporate seasonality through additional seasonal orders $(P,D,Q)$ and seasonal frequency of the data, $m$ \citep{hyn:ath:21:3ed}.

In this study, ARIMA models are fitted in R using the \texttt{auto.arima()} function from the \texttt{forecast} package. The function automatically selects an ARIMA specification from the candidate models using the corrected Akaike information criterion (AICc) under its default settings. Assuming Gaussian forecast errors, the corresponding predictive quantiles are obtained using Eq.~\eqref{eq:ETS:PI}. The resulting forecast-error standard deviations account for uncertainty due to future innovations, but do not include uncertainty arising from estimation of the ARIMA coefficients.

\subsection{Forecast evaluation}
\label{ssec:Forecast:evaluation}

We evaluate the probabilistic forecast accuracy of the post-processing approaches described in Section~\ref{sec:post-processing} relative to the benchmark predictive distributions provided by the \texttt{forecast} package in R, as described in Section~\ref{ssec:Point:Models}. As the evaluation metric, we use the \textit{continuous ranked probability score} \cite[CRPS;][]{gne:raf:07}, a strictly proper scoring rule for predictive distributions. In integral form, the CRPS is defined as
\begin{equation}
\text{CRPS}\left(\widehat{F},x\right) = \int_{-\infty}^{\infty} \left( \widehat{F}(y)-\mathbf{1}_{\{x\leq y\}} \right)^2 \,dy,
\label{eq:CRPS}
\end{equation}
where $\widehat{F}$ is the predictive cumulative distribution function, $x$ is the observed value, and $\mathbf{1}_{\{x\leq y\}}$ is the indicator function.

In practice, we approximate the CRPS using a finite set of quantile forecasts:
\begin{equation}
\text{CRPS}\left(\widehat{F},x\right) \approx \frac{2}{M} \sum_{i=1}^{M} \text{PS}\left(\hat{q}^{p_i},x,p_i\right),
\label{eq:CRPS:approx}
\end{equation}
where $\hat{q}^{p_i}\equiv\widehat{F}^{-1}(p_i)$ is the forecast of the $p_i$-quantile, $(p_1,\ldots,p_M)$ is an equally spaced grid of quantile levels, and $\text{PS}\left(\hat{q}^{p},x,p\right)$ denotes the \textit{pinball score},
\begin{equation}
\text{PS}\left(\hat{q}^{p},x,p\right) = \left( \mathbf{1}_{\{x<\hat{q}^{p}\}}-p \right) \left(\hat{q}^{p}-x \right),
\label{eq:PS}
\end{equation}
also known as the \textit{pinball loss}, \textit{quantile loss}, or \textit{check function} \citep{ber:zie:23,g-c:etal:17,now:wer:18}. Following common practice in probabilistic forecasting literature and competitions \citep{hon:pin:fan:etal:16}, we use $M=99$ quantile levels, $p\in\{0.01,0.02,\ldots,0.99\}$. The pinball score is asymmetric for $p\neq0.5$, penalizing underprediction and overprediction differently depending on the quantile level.

Since the M4 monthly series come from different domains and vary substantially in scale, raw CRPS values are not directly comparable across series. We therefore evaluate each post-processing approach relative to the corresponding benchmark using the relative CRPS:
\begin{equation}
\label{eq:rCRPS}
\text{rCRPS}_{h,i} = \frac{ \sum_{\tau\in\mathcal{T} {\text{test}}} \text{CRPS}_{h,i,\tau}^{(\text{model})} }{ \sum_{\tau\in\mathcal{T}_{\text{test}}} \text{CRPS}_{h,i,\tau}^{(\text{benchmark})} },
\end{equation}
where $h=1,\ldots,12$ denotes the forecast horizon, $i=1,\ldots,14{,}407$ indexes the time series, and $\tau\in\mathcal{T}_{\text{test}}$ denotes the forecast target, as defined in Section~\ref{ssec:Notation}. Clearly, $\text{rCRPS}_{h,i}<1$ indicates that the post-processing approach outperforms the benchmark, whereas $\text{rCRPS}_{h,i}>1$ indicates worse performance.

To assess overall forecast accuracy at each horizon $h$, we aggregate $\text{rCRPS}_{h,i}$ across series using the geometric mean, computed by exponentiating the arithmetic mean of $\ln(\text{rCRPS}_{h,i})$. This gives equal multiplicative weight to relative improvements and performance losses with respect to the benchmark. We express the resulting measure as the \textit{continuous ranked probability skill score}:
\begin{equation}
\label{eq:skillscore}
\text{CRPSS}_{h} = \left[ 1- \exp\left( \frac{1}{N} \sum_{i=1}^{N} \ln\left(\text{rCRPS}_{h,i}\right) \right) \right] \times 100\%.
\end{equation}
Thus, $\text{CRPSS}_{h}>0$ indicates an improvement over the benchmark, whereas $\text{CRPSS}_{h}<0$ indicates worse performance.

\section{Results}
\label{sec:Results}

\subsection{Probabilistic forecasting performance}
\label{sec:CRPSresults}

Table~\ref{tab:relcrps_results} summarizes the CRPSS values for the three base models, four post-processing methods, and two calibration approaches, averaged across forecast horizons $h=1,\ldots,12$. All reported CRPSS values are positive, indicating that, on average across series and horizons, every post-processing variant improves upon the corresponding benchmark predictive distribution.

\begin{table}[tb]
\centering
\caption{Continuous ranked probability skill scores (CRPSS), as defined in Eq.~\eqref{eq:skillscore}, for the three forecasting models (Theta, ETS, and ARIMA; \textit{columns}) and the in-sample and out-of-sample variants of four post-processing schemes (HS, CP, QR, and GARCH; \textit{rows}).}
\label{tab:relcrps_results}
\resizebox{\textwidth}{!}{%
\begin{tabular}{|l|ccc|ccc|ccc|}
\hline
 &
  \multicolumn{3}{c|}{\textbf{Theta}} &
  \multicolumn{3}{c|}{\textbf{ETS}} &
  \multicolumn{3}{c|}{\textbf{ARIMA}} \\ \hline
 &
  \multicolumn{1}{c|}{In-sample} &
  \multicolumn{1}{c|}{Out-of-sample} &
  Difference &
  \multicolumn{1}{c|}{In-sample} &
  \multicolumn{1}{c|}{Out-of-sample} &
  Difference &
  \multicolumn{1}{c|}{In-sample} &
  \multicolumn{1}{c|}{Out-of-sample} &
  Difference \\ \hline
\textbf{CP} &
  \multicolumn{1}{c|}{\cellcolor[HTML]{71D39E}2.56\%} &
  \multicolumn{1}{c|}{\cellcolor[HTML]{7ED7A7}2.33\%} &
  \text{0.23\%} &
  \multicolumn{1}{c|}{\cellcolor[HTML]{09B356}3.14\%} &
  \multicolumn{1}{c|}{\cellcolor[HTML]{73D49F}1.79\%} &
  \text{1.35\%} &
  \multicolumn{1}{c|}{\cellcolor[HTML]{31BF72}3.67\%} &
  \multicolumn{1}{c|}{\cellcolor[HTML]{42C57D}3.37\%} &
  \text{0.30\%} \\ \hline
\textbf{HS} &
  \multicolumn{1}{c|}{\cellcolor[HTML]{53CA89}3.10\%} &
  \multicolumn{1}{c|}{\cellcolor[HTML]{70D39D}2.58\%} &
  \text{0.52\%} &
  \multicolumn{1}{c|}{\cellcolor[HTML]{00B050}3.25\%} &
  \multicolumn{1}{c|}{\cellcolor[HTML]{75D5A1}1.76\%} &
  \text{1.49\%} &
  \multicolumn{1}{c|}{\cellcolor[HTML]{39C278}3.52\%} &
  \multicolumn{1}{c|}{\cellcolor[HTML]{4EC986}3.15\%} &
  \text{0.37\%} \\ \hline
\textbf{QR} &
  \multicolumn{1}{c|}{\cellcolor[HTML]{00B050}4.59\%} &
  \multicolumn{1}{c|}{\cellcolor[HTML]{3EC47B}3.48\%} &
  \text{1.11\%} &
  \multicolumn{1}{c|}{\cellcolor[HTML]{D7F3E4}0.52\%} &
  \multicolumn{1}{c|}{\cellcolor[HTML]{4BC884}2.30\%} &
  \text{$-$1.78\%} &
  \multicolumn{1}{c|}{\cellcolor[HTML]{00B050}4.53\%} &
  \multicolumn{1}{c|}{\cellcolor[HTML]{23BB68}3.91\%} &
  \text{0.62\%} \\ \hline
\textbf{GARCH} &
  \multicolumn{1}{c|}{\cellcolor[HTML]{8CDCB0}2.08\%} &
  \multicolumn{1}{c|}{\cellcolor[HTML]{C0ECD4}1.15\%} &
  \text{0.93\%} &
  \multicolumn{1}{c|}{\cellcolor[HTML]{EFFAF4}0.21\%} &
  \multicolumn{1}{c|}{\cellcolor[HTML]{FEFFFE}0.02\%} &
  \text{0.19\%} &
  \multicolumn{1}{c|}{\cellcolor[HTML]{69D198}2.67\%} &
  \multicolumn{1}{c|}{\cellcolor[HTML]{9BE0BB}1.78\%} &
  \text{0.89\%} \\ \hline
\end{tabular}%
}
\end{table}

The magnitude of the improvement depends strongly on both the base model and the post-processing method. For Theta forecasts, $\text{QR}_{\text{in}}$ performs best, with a CRPSS of $4.59\%$, followed by $\text{QR}_{\text{out}}$ at $3.48\%$. A similar pattern is observed for ARIMA, for which $\text{QR}_{\text{in}}$ and $\text{QR}_{\text{out}}$ achieve CRPSS values of $4.53\%$ and $3.91\%$, respectively. For ETS forecasts, however, HS and CP perform best: $\text{HS}_{\text{in}}$ achieves a CRPSS of $3.25\%$, closely followed by $\text{CP}_{\text{in}}$ at $3.14\%$. In contrast, $\text{QR}_{\text{in}}$, which performs best for Theta and ARIMA, yields only a $0.52\%$ improvement for ETS. These results indicate that no post-processing method dominates independently of the underlying forecasting model.

A particularly notable result is the comparison between in-sample and out-of-sample calibration. The in-sample variant performs better in 11 of the 12 model--method combinations considered. The largest advantages occur for HS and CP applied to ETS forecasts, for which the CRPSS differences are $1.49$ and $1.35$ percentage points, respectively. The only exception is QR applied to ETS, for which the out-of-sample variant outperforms the in-sample variant by $1.78$ percentage points.

\begin{figure}[tb]
    \centering
    \includegraphics[width=\linewidth]{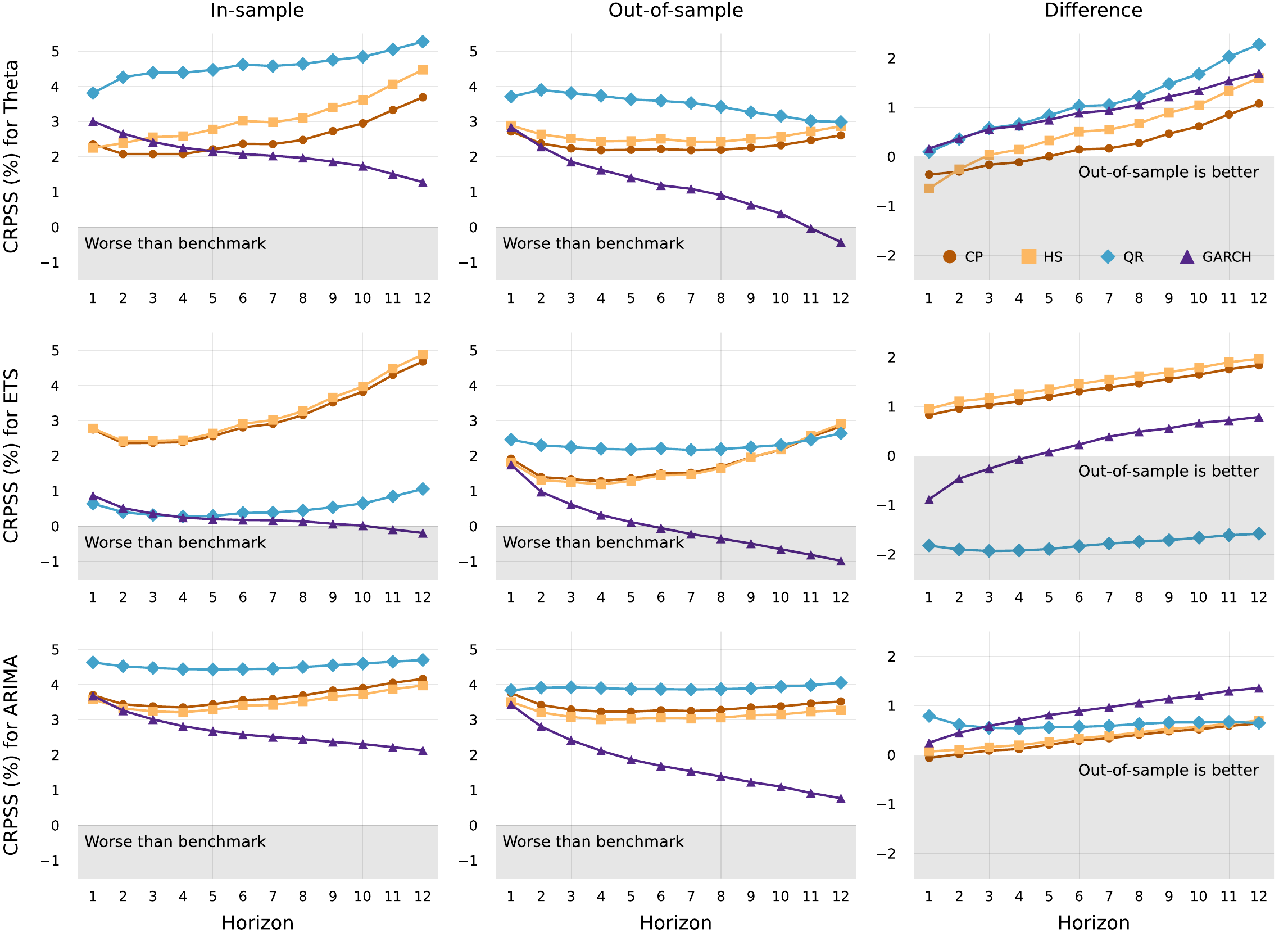}
    \caption{CRPSS, as defined in Eq.~\eqref{eq:skillscore}, for each forecast horizon $h=1,\ldots,12$, forecasting model (Theta, ETS, and ARIMA; \textit{rows}), and post-processing setting. The left and middle columns show the results for in-sample and out-of-sample post-processing, respectively. The right column shows the difference between the two.}
    \label{fig:CRPSS}
\end{figure}

Figure~\ref{fig:CRPSS} shows that these aggregate results mask substantial variation across forecast horizons. For most combinations, the relative advantage of in-sample calibration increases with the horizon. This pattern is particularly pronounced for Theta and for HS and CP applied to ETS. QR applied to ETS is the main exception, with out-of-sample calibration remaining superior throughout the horizon range.

GARCH exhibits a different horizon profile. Its CRPSS generally decreases as the forecast horizon increases, and some GARCH variants eventually perform worse than the benchmark at longer horizons. Thus, although all GARCH variants have positive CRPSS values when averaged across horizons, their relative over-performance is concentrated primarily at shorter horizons.

\subsection{Ranked performance analysis}
\label{sec:MCBresults}

To complement the CRPSS analysis, we apply multiple comparisons with the best \citep[MCB;][]{Koning2005-iq} tests to assess whether differences in performance are statistically significant in terms of ranks across series. For the aggregate analysis, CRPS values are first averaged across horizons $h=1,\ldots,12$ for each series. Separate MCB tests are then conducted for Theta, ETS, and ARIMA forecasts. Horizon-specific results for $h=1,3,6,9,$ and $12$ are reported in \ref{sec:Appendix}.

\begin{figure}[tb]
    \centering
    \includegraphics[width=1\linewidth]{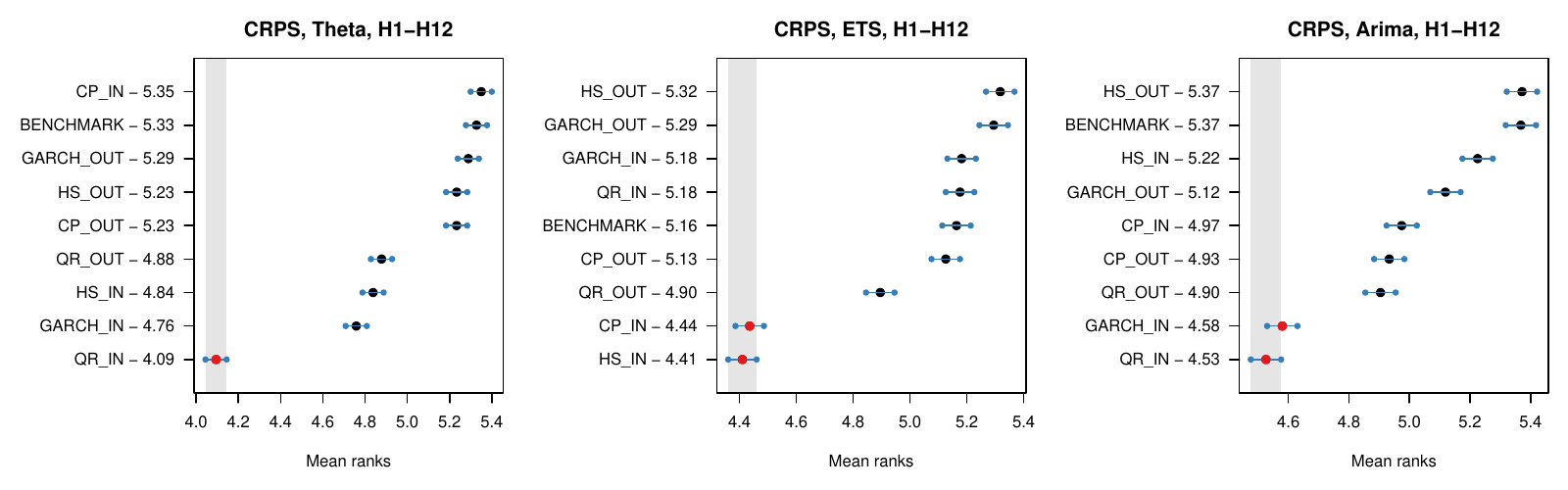}
    \caption{MCB test results for the CRPS of all considered post-processing schemes, aggregated across all forecast horizons $h=1,\ldots,12$, for forecasts generated by Theta, ETS, and ARIMA (\textit{left to right}). For each approach, we report the mean rank and its rank interval; non-overlapping intervals indicate statistically significant differences in ranks. Approaches whose intervals overlap the shaded region around the best-ranked method are statistically indistinguishable from it.}
    \label{fig:MCB_results_In_Out}
\end{figure}

Figure~\ref{fig:MCB_results_In_Out} confirms that the preferred post-processing approach depends on the base model. For Theta, $\text{QR}_{\text{in}}$ has the lowest mean rank and clearly dominates the aggregate ranking. For ETS, $\text{HS}_{\text{in}}$ and $\text{CP}_{\text{in}}$ have the two lowest mean ranks and are statistically indistinguishable from one another. 
For ARIMA, $\text{QR}_{\text{in}}$ and $\text{GARCH}_{\text{in}}$ achieve the best ranks and are statistically indistinguishable within the rank-based approach.

Two methods, $\text{HS}_{\text{in}}$ and $\text{QR}_{\text{out}}$, significantly outperform the benchmark for all three base models. The comparison with Table~\ref{tab:relcrps_results} also highlights the distinction between the magnitude and consistency of forecast improvements. A method can achieve a positive aggregate CRPSS while not ranking consistently above the benchmark across individual series. Thus, the CRPSS and MCB analyses provide complementary evidence on average gains and cross-series robustness.

The horizon-specific MCB results reinforce the patterns observed in Figure \ref{fig:CRPSS}. For Theta, $\text{QR}_{\text{in}}$ remains the best-ranked method for all 12 horizons. For ETS, the best-ranked method changes from $\text{GARCH}_{\text{out}}$ at $h=1$ and $\text{QR}_{\text{out}}$ at $h=3$ to $\text{CP}_{\text{in}}$ at $h=6$ and $\text{HS}_{\text{in}}$ at $h=9$ and $12$. For ARIMA, $\text{GARCH}_{\text{in}}$ performs particularly strongly at short horizons, whereas $\text{QR}_{\text{in}}$ becomes the leading method at medium and longer horizons. Overall, these results provide further evidence that both the preferred post-processing method and the relative value of in-sample calibration depend on the forecast horizon.

\subsection{Computational cost}

In this section, we measure the computational cost of the benchmark and the in- and out-of-sample post-processing approaches. We consider two components: (i) the cost of generating the forecasts and (ii) the cost of post-processing the corresponding prediction errors. In detail:
\begin{itemize}[noitemsep]
\item the benchmark incurs only the cost of producing quantile forecasts using the $T-K$ observations available at the forecast origin;
\item the in-sample approaches incur the cost of generating forecasts using the same $T-K$ observations as the benchmark and the cost of post-processing the in-sample prediction errors; 
\item the out-of-sample approaches incur the cost of producing rolling-origin forecasts for the calibration window and the cost of post-processing the out-of-sample prediction errors.
\end{itemize}

Table \ref{tab:cost} reports the computational costs averaged across a sample of 100 series. The results are presented separately for each forecasting model (Theta, ETS, and ARIMA) and each post-processing approach (CP, HS, QR, and GARCH). The computational cost of generating forecasts differs substantially across the three forecasting models. Theta is by far the fastest, while ETS and ARIMA are about 50 and 130 times slower, respectively. Generating the rolling-origin forecasts required for out-of-sample calibration is considerably more demanding. It increases the computational times by approximately 180-215 times relative to the corresponding benchmark/in-sample setting.

\begin{table}[tb]
\caption{Computation time per series at forecast origin $\xi=312$, estimated based on 100 series and averaged using the arithmetic mean. The upper panel reports the time required to generate forecasts for the benchmark/in-sample setting and the rolling-origin forecasts required for out-of-sample calibration. The lower panel reports the additional computation time required by each post-processing method. Base forecasts were generated in R~4.5.3 and post-processing was performed in Julia~1.12.6, using a single execution thread on an Apple M2 Pro.}
\label{tab:cost}
\center
\begin{tabular}{|l|c|c|}
\hline
\textbf{Forecasting model} & \textbf{In-sample/benchmark}   & \textbf{Out-of-sample} \\
\hline
Theta & 8.4~ms & 1.8~s  \\
ETS   & 0.45~s & 83~s   \\
ARIMA & 1.1~s & 3.3~m  \\
\hline
\textbf{Post-processing approach} & \textbf{In-sample}   & \textbf{Out-of-sample} \\
\hline
CP    & 0.20~ms & 0.16~ms    \\
HS    & 0.20~ms & 0.16~ms   \\
QR    & 0.12~s & 0.10~s   \\
GARCH & 2.2~ms & 1.1~ms \\
\hline
\end{tabular}
\end{table}

On the other hand, the computational cost for post-processing the prediction errors is generally small. For example, in-sample CP and HS require only 0.20~ms per series, corresponding to an additional computational cost of about $2.5\%$ relative to generating Theta forecasts, and considerably less relative to ETS and ARIMA. In-sample GARCH is about 11 times slower than CP and HS, while in-sample QR is by far the most computationally intensive post-processing approach (about 600 slower than in-sample CP/HS). Still, such post-processing times are relatively small compared to the cost of producing the base forecasts, especially for ETS and ARIMA.

Comparing the computational cost of the post-processing step itself, we observe that the out-of-sample variants are generally faster than their in-sample counterparts. The reduction ranges from about $17\%$ for QR to $50\%$ for GARCH, reflecting the shorter calibration samples used in out-of-sample post-processing. However, the lower cost of out-of-sample post-processing cannot compensate for the additional cost of generating out-of-sample forecasts.

\section{Discussion}
\label{sec:Discussion}

This study addresses two related questions: (\textit{i}) whether post-processing point-prediction errors improves upon the default predictive distributions of standard forecasting models, and (\textit{ii}) whether post-processing is more effective when calibrated using in-sample residuals or out-of-sample forecast errors.
The empirical results provide clear evidence on both questions. When averaged across forecast horizons, all post-processing variants considered in this study improve upon their corresponding benchmark predictive distributions. Moreover, as reported in Table \ref{tab:relcrps_results}, the in-sample variant achieves a higher average CRPSS than its out-of-sample counterpart in 11 of the 12 model--method comparisons.

The magnitude of the gains depends substantially on the underlying forecasting model and post-processing method. QR performs particularly well when applied to Theta and ARIMA forecasts, whereas HS and CP are more effective for ETS. This heterogeneity indicates that post-processing should not be viewed as a universally interchangeable final step. Rather, its effectiveness depends on the characteristics of the errors generated by the underlying forecasting model. In business applications, the choice of post-processing method should therefore be considered jointly with the base forecasting model.

The comparison between in-sample and out-of-sample calibration provides one of the main findings of the study. Although out-of-sample forecast errors are often regarded as more representative of genuine forecasting performance, their use does not systematically translate into more accurate probabilistic forecasts in our empirical setting. In-sample calibration performs better in almost all direct comparisons, with QR applied to ETS forecasts being the only exception in Table \ref{tab:relcrps_results}. This result suggests that readily available in-sample residuals can provide a useful basis for post-processing, despite being obtained from observations that were also used for model training.

As illustrated in Figure \ref{fig:CRPSS}, the relative performance of the methods also varies with the forecast horizon. For many combinations, the advantage of in-sample calibration increases as the horizon becomes longer. This pattern is especially visible for Theta and for HS and CP applied to ETS forecasts. GARCH exhibits a different profile: its relative performance is strongest at shorter horizons and generally declines as the horizon increases, in some cases falling below the benchmark. These results suggest that different post-processing approaches capture different features of forecast uncertainty and that their relative suitability may change as uncertainty accumulates over the forecast horizon.

The rank-based analysis presented in Figure \ref{fig:MCB_results_In_Out} and in \ref{sec:Appendix} complements the CRPSS results. CRPSS measures the magnitude of the improvement relative to the benchmark, whereas the MCB analysis emphasizes how consistently a method performs across individual series. Consequently, a method can achieve a positive aggregate CRPSS without obtaining a substantially better mean rank than the benchmark. Conversely, some methods with more moderate average improvements perform consistently well across a broad range of series. The two measures therefore provide complementary perspectives on probabilistic forecast performance: one captures the size of the gain, while the other captures its cross-series robustness.

From a managerial perspective, these findings are relevant for organisations that already rely on established point-forecasting systems but require uncertainty information for operational decisions. Post-processing provides a way to enhance such systems without replacing the underlying forecasting models. This can be useful in applications such as inventory management, workforce scheduling, budgeting, and capacity planning, where decisions depend not only on expected outcomes but also on the probability of unusually high or low realizations.

Computational cost provides an additional argument in favor of in-sample calibration when historical forecasts are not already available. As shown in Table~\ref{tab:cost}, the additional cost of post-processing itself is generally small relative to the cost of generating the base forecasts. The main computational difference arises from the need to reconstruct the historical one-step-ahead forecasts used for out-of-sample calibration. This increases forecast generation time by approximately 200 times, with the absolute difference becoming particularly large for ARIMA. In-sample calibration therefore offers an attractive combination of forecast accuracy and computational efficiency: it performs better in 11 of the 12 model--method comparisons while avoiding repeated historical model estimation.

This advantage depends, however, on the information infrastructure available to the forecaster. If historical point forecasts and their realizations are routinely stored, the forecast-generation cost associated with rolling-origin calibration has already been incurred, and only the comparatively small post-processing cost remains. The computational advantage of in-sample calibration is therefore greatest when probabilistic forecasts are being added retrospectively to an existing point-forecasting system.

An important limitation concerns the construction of the calibration samples. In our design, the out-of-sample calibration sample contains 72 fewer forecast--observation pairs than the corresponding in-sample calibration sample. The observed performance differences may therefore reflect not only the source of the calibration errors but also the amount of information available for post-processing. Nevertheless, given the available 312 monthly observations, this design provides the fairest comparison between the in-sample and out-of-sample approaches, as both make use of all information available at each forecast origin.

A second limitation concerns the horizon-specific scale adjustment. Both calibration approaches rely on model-based estimates of how forecast uncertainty changes with the horizon. The performance observed at longer horizons therefore reflects the combination of the post-processing method and the underlying horizon-scaling mechanism. Future research could investigate alternative scaling procedures and assess how sensitive the conclusions are to this component of the framework.

Finally, previous research has demonstrated the effectiveness of post-processing methods in high-frequency settings, where long historical records are typically available. In this study, we focus on monthly time series, which represent a different forecasting context and business setting. Extending the analysis to other data frequencies would therefore be valuable, including intermediate frequencies such as weekly data and lower frequencies such as quarterly or yearly data. We expect the benefits of post-processing to diminish as data frequency decreases, because fewer observations may limit the information available for reliable error calibration. Examining the relationship between data frequency, calibration-sample size, and probabilistic forecast performance remains an important direction for future research.

\section{Conclusions}
\label{sec:Conclusions}

This study examined whether probabilistic forecasts obtained by post-processing prediction errors can improve upon the default predictive distributions of widely used forecasting models and whether post-processing should be based on in-sample residuals or out-of-sample forecast errors. We considered four post-processing methods (historical simulation, conformal prediction, quantile regression, and GARCH) applied to point forecasts generated by three classical time series models (Theta, exponential smoothing, and ARIMA) and evaluated them on 14,407 monthly series from the M4 forecasting competition.

The results show that post-processing is effective: when averaged across forecast horizons, all considered variants improve upon their corresponding benchmark predictive distributions. In-sample calibration performs better than its out-of-sample counterpart in 11 of the 12 model--method comparisons. At the same time, no single post-processing method dominates across all settings: QR performs particularly well for Theta and ARIMA, whereas HS and CP are more effective for ETS, and relative performance varies with the forecast horizon. 
The additional cost of post-processing is generally small relative to model estimation, whereas reconstructing the rolling-origin forecasts required for out-of-sample calibration can be computationally expensive.

From a practical perspective, these findings indicate that existing point-forecasting systems can be extended with useful uncertainty information without necessarily changing the underlying forecasting models. In many settings, readily available in-sample residuals provide an effective basis for probabilistic forecast post-processing. Future research could further examine the role of calibration-sample size, alternative horizon-scaling procedures, and other data frequencies.

\section*{CRediT} 

Conceptualization -- FP, RW; Data curation -- FP; Funding acquisition -- RW; Investigation -- AL, FP, PZ; Methodology -- AL, FP, RW, PZ; Software -- AL, FP, PZ; Supervision -- RW; Validation -- AL, FP, RW; Visualization -- AL, PZ; Writing (original draft) -- AL, FP, PZ, RW; Writing (review \& editing) -- FP, RW.

\section*{Acknowledgments}

The study was partially supported by the National Science Centre (NCN, Poland) through grant no.\ 2025/57/B/HS4/02413 (to PZ) and by the National Science Centre (NCN, Poland) and the German Research Foundation (DFG, Germany) through grant no. 2021/43/I/HS4/02578 (to AL and RW; DFG no. 505565850).

{\small
\setlength{\bibsep}{0pt}
\bibliographystyle{elsarticle-harv} 
\bibliography{epf}
}

\newpage

\appendix
\section{MCB test results per forecast horizon}
\label{sec:Appendix}

\begin{figure}[H]
    \centering
    \includegraphics[width=0.85\linewidth]{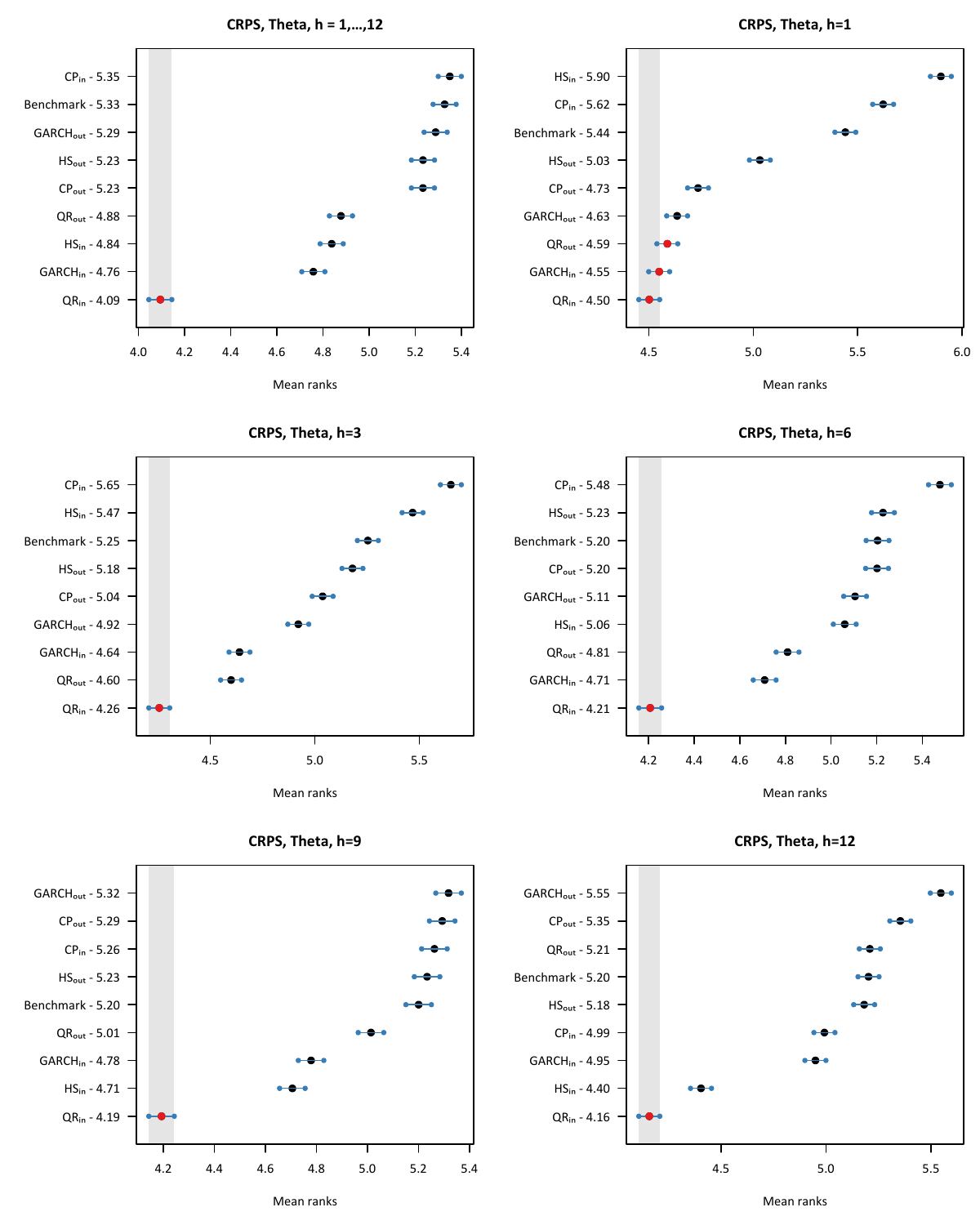}
    \caption{MCB test results per horizon for CRPS and Theta forecasts. Note the dominance of quantile regression in-sample (QR$_\text{in}$) post-processing nearly across all horizons.}
    \label{figapp:MCB_RelCRPS_THETA}
\end{figure}

\begin{figure}[p]
    \centering
    \includegraphics[width=0.85\linewidth]{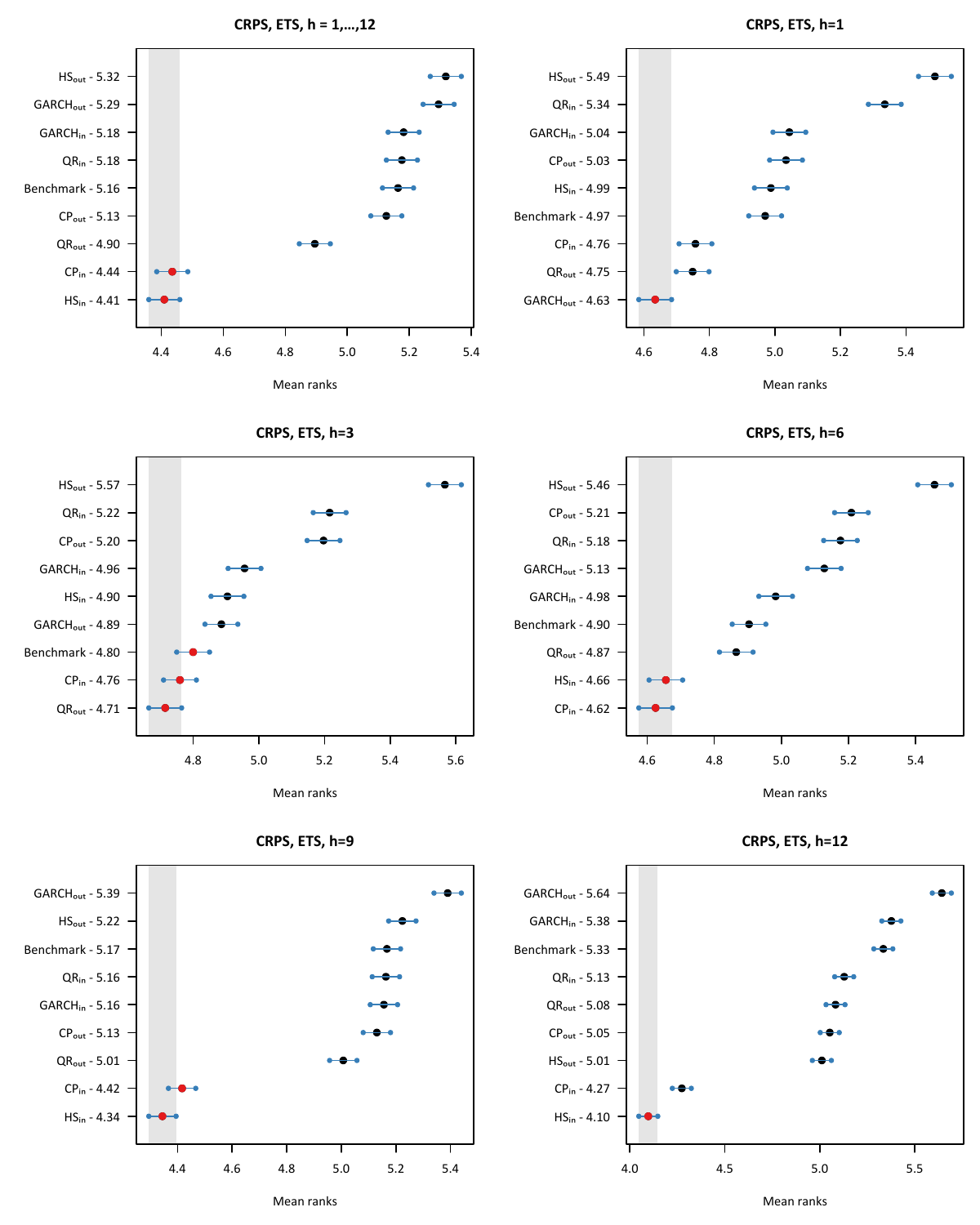}
    \caption{MCB test results per horizon for CRPS and ETS forecasts.}
    \label{figapp:MCB_RelCRPS_ETS}
\end{figure}

\begin{figure}[p]
    \centering
    \includegraphics[width=0.85\linewidth]{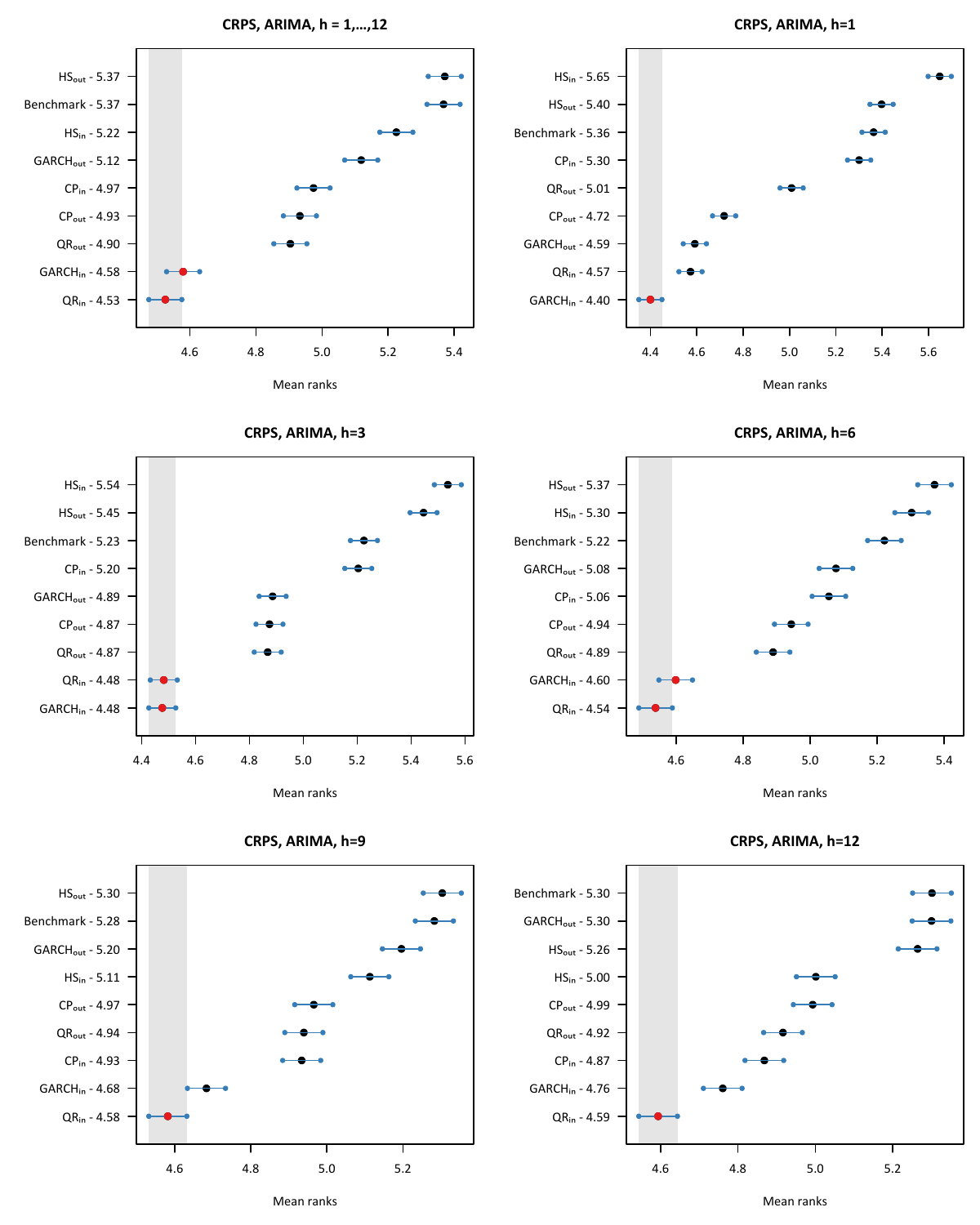}
    \caption{MCB test results per horizon for CRPS and ARIMA forecasts.}
    \label{figapp:MCB_RelCRPS_ARIMA}
\end{figure}

This appendix complements the aggregate MCB analysis presented in Section~\ref{sec:Results} by reporting results for selected forecast horizons. Figures~\ref{figapp:MCB_RelCRPS_THETA}--\ref{figapp:MCB_RelCRPS_ARIMA} present the MCB results for $h=1,3,6,9,$ and $12$, separately for Theta, ETS, and ARIMA forecasts. 
\color{black}

\end{document}